\documentstyle[aps]{revtex}
\begin{document}
\input epsf
\draft \tighten 
 \twocolumn[\hsize\textwidth\columnwidth\hsize\csname
 @twocolumnfalse\endcsname \preprint{CERN-TH/99-296, hep-ph/9909508}
\title{Gravitational Particle Production and the Moduli Problem}
\author{Gary Felder$^{1,2}$,~ Lev Kofman$^{2}$,~ and Andrei
Linde$^{1,3}$} 
\address{$^{1}$Department of Physics, Stanford University, Stanford,
CA 94305, USA} 
\address{$^{2}$CITA, University of Toronto, 60 St George Str,
Toronto, ON M5S 3H8, Canada} 
\address{$^{3}$Theory Division, CERN CH-1211, Geneva 23, Switzerland}
\date{September 24, 1999}
\maketitle 
\begin{abstract}
A theory of gravitational production of light scalar particles during and 
after inflation is investigated. We show that in the most interesting 
cases where long-wavelength fluctuations of light scalar fields can be 
generated during inflation, these fluctuations rather than quantum 
fluctuations produced after inflation  give the dominant contribution to  
particle production. In such cases a simple analytical theory of particle 
production can be developed. Application of our results to the theory of 
quantum creation of moduli fields demonstrates that if the moduli mass is 
smaller than the Hubble constant then these fields are copiously produced 
during inflation. This gives rise to the cosmological moduli problem even 
if there is no homogeneous component of the classical moduli field in the 
universe. To avoid this version of the moduli problem it is necessary for 
the Hubble constant $H$ during the last stages of inflation and/or the 
reheating temperature $T_R$ after inflation to be extremely small. 
\end{abstract}
\pacs{PACS: 98.80.Cq  \hskip 1.5cm CERN-TH/99-296 \hskip 1.5cm CITA-99-30 
\hskip 1.5cm hep-ph/9909508} 
 \vskip2pc]

\section{Introduction}
Recently there has been a renewal of interest in gravitational production 
of particles in an expanding universe. This   was a subject of intensive 
study many years ago, see e.g. \cite{grib}. However, with the invention 
of inflationary theory the issue of the production of particles due to 
gravitational effects became less urgent. Indeed, gravitational effects 
are especially important near the cosmological singularity, at the Planck 
time. But the density of the particles produced at that epoch becomes 
exponentially small due to inflation. New particles are produced only 
after the end of inflation when the energy density is much smaller than 
the Planck density. Production of particles due to gravitational effects 
at that stage is typically very inefficient.

There are a few exceptions to this rule that have motivated the recent 
interest in gravitational particle production. First of all, there are 
some models where the main mechanism of reheating during inflation is due 
to gravitational production. Even though this mechanism is very 
inefficient, in the absence of other mechanisms of reheating it may do 
the job. For example, one may consider the class of theories where the 
inflaton potential $V(\phi)$ gradually decreases at large $\phi$ and does 
not have any minima. In such theories the inflaton field $\phi$ does not 
oscillate after inflation,  so the standard mechanism of reheating does 
not work \cite{PV,NO}. To emphasize this unusual feature of such theories 
we call them  non-oscillatory models, or simply NO models \cite{NO}. 
Usually gravitational particle production in such models lead to 
dangerous cosmological consequences, such as large isocurvature 
fluctuations and overproduction of gravitinos \cite{NO}. In order to 
overcome these problems, it was necessary to modify the NO models 
\cite{NO} and to use the non-gravitational mechanism of instant 
preheating for the description of particle production \cite{inst}. 

There are some other cases where even very small but unavoidable 
gravitational particle production may lead either to useful   or to 
catastrophic consequences \cite{ch,tk,GLV,KKLV,GRT}. For example, it has 
recently been found that the production of gravitinos by the oscillating 
inflaton field is not suppressed by the small gravitational coupling. As 
a result, gravitinos can be copiously produced in the early universe even 
if the reheating temperature always remains smaller than $10^8$ GeV 
\cite{KKLV,GRT}. Another important example is related to moduli 
production. 15 years ago Coughlan {\it et al} realized that string theory 
and supergravity give rise to a cosmological moduli problem associated 
with the existence of a large homogeneous classical moduli field in the 
early universe \cite{polonyi}. Soon afterwards Goncharov, Linde and 
Vysotsky showed that quantum fluctuations of moduli fields produced at 
the last stages of inflation  lead  to the moduli problem even if 
initially there were no  classical moduli fields \cite{GLV}. 
 Thus the cosmological moduli problem may appear either because of
the existence of a large  long-living  homogeneous classical moduli field 
or because  of   quantum production of excitations (particles) of the 
moduli fields. In \cite{NO} it was pointed out that the problem of moduli 
production is especially difficult in the context of NO models, where 
moduli are produced as abundantly as usual particles. 

Recently the problem of  moduli production in the early universe was 
studied by numerical methods in \cite{GRT}, with conclusions similar to 
those of Ref. \cite{GLV}.   As we are going to demonstrate, the main 
source of gravitational production of light moduli  in inflationary 
cosmology  is very simple,  and  one can study the theory of moduli 
production not only numerically but also analytically by the methods 
developed in \cite{GLV,NO}. This will allow us to generalize and 
considerably strengthen  the results of Refs. \cite{GLV,GRT}. 

In particular, we will see that in the leading approximation the problem 
of overproduction of light moduli particles is {\it equivalent}  to the 
problem of large homogeneous classical moduli fields \cite{GLV}. We will 
show that the ratio of the number density of light moduli produced during 
inflation to the entropy of the universe after reheating satisfies the 
inequality 
\begin{equation}\label{pot4}
{n_\chi\over s} \gtrsim   { T_R H_0^2\over 3 m M_p^2} \ . 
\end{equation}
Here $m$ is the moduli mass, $M_p \sim 2.4 \times 10^{18}$ GeV is the 
reduced Planck mass, and $H_0$ is the Hubble constant at the moment 
corresponding to the beginning of the last 60 e-foldings before the end 
of inflation. 

In the simplest versions of inflationary theory with potentials 
$M^2\phi^2/2$ or $\lambda\phi^4/4$ one has  $H_0 \sim 10^{14}$ GeV. In 
such models our result implies that in order to satisfy the cosmological 
constraint ${n_\chi\over s} \lesssim 10^{-12}$ one needs to have an 
abnormally small reheating temperature $T_R \lesssim 1$ GeV. 
Alternatively one may consider inflationary models where  the Hubble 
constant at the end of inflation is very small. But we will argue that 
even this may not help, so one may need either to invoke  thermal 
inflation or to use some other mechanisms which can make the moduli 
problem less severe, see e.g. \cite{LS,LinMod}. 

 In the next section we outline the classical and quantum
versions of the moduli problem and explain how each of them can arise in 
inflationary theory. In section III we describe the results of our 
numerical simulations of gravitational production of light scalar fields 
during and after preheating. In particular we verify our prediction that 
the dominant contribution to particle production comes from 
long-wavelength modes which are indistinguishable from homogeneous 
classical moduli fields. Finally in section IV we analytically compute 
the production of these long wavelength modes and derive Eq.(\ref{pot4}). 
This section also contains our concluding discussion.

\section{Moduli problem}

String moduli  couple to standard model fields only through Planck scale 
suppressed interactions. Their effective potential is exactly flat  in 
perturbation theory  in the supersymmetric limit, but it may become 
curved due to nonperturbative effects or because of supersymmetry 
breaking. If these fields originally are far from the minimum of their 
effective potential, the energy of their oscillations will decrease in an 
expanding universe in the same way as the energy density of 
nonrelativistic matter, $\rho_m \sim a^{-3}(t)$. Meanwhile the energy 
density of relativistic plasma decreases as $a^{-4}$. Therefore the 
relative contribution of moduli to the energy density of the universe may 
quickly become very significant. They are expected to decay after the 
stage of nucleosynthesis, violating the standard nucleosynthesis 
predictions unless the initial amplitude of the moduli oscillations 
$\chi_0$ is sufficiently small. The constraints  on the energy density of 
the moduli field $\rho_\chi$ and the number of moduli particles $n_\chi$ 
depend  on details of the theory. The most stringent constraint appears 
because of the photodissociation and photoproduction of light elements by 
the decay products of the moduli fields. For $m \sim 10^2 - 10^3$ GeV one 
has 
\begin{equation}\label{constr}
{n_\chi\over s} \lesssim  10^{-12}-10^{-15} \ . 
\end{equation}
 see   \cite{constraint,rt} and references therein.
In this paper we will use  a conservative version of this constraint, 
 ${n_\chi\over s} \lesssim   10^{-12}$.

If the field $\chi$ is a classical homogeneous oscillating scalar field, 
then this constraint applies to it as well if one defines the 
corresponding number density of nonrelativistic particles $\chi$ by the 
following obvious relation: 
\begin{equation}\label{constra}
{n_\chi } = {\rho_\chi\over m} = {m\chi^2\over 2} \ . 
\end{equation}

Let us first consider  moduli $\chi$ with a constant mass $m \sim 10^2 - 
10^3$ GeV and assume that reheating of the universe  occurs after the 
beginning of oscillations of the moduli. This  is indeed the case if one 
takes into account that in order to avoid the gravitino problem one 
should have $T_R < 10^8$ GeV. We will also assume for definiteness that 
the minimum of the effective potential for the field $\chi$ is at $\chi = 
0$; one can always achieve this by an obvious redefinition of the field 
$\chi$.

Independent of the choice of inflationary theory, at the end of inflation 
the main fraction of the energy density of the universe is concentrated 
in the energy of an oscillating scalar field $\phi$. Typically this is 
the same  field  which drives inflation, but in some models such as 
hybrid inflation this may not be the case. We will consider here the 
simplest (and most general) model where the effective potential of the  
field $\phi$ after inflation is quadratic, 
\begin{equation}\label{pot}
V(\phi) = {M^2\over 2}\phi^2. 
\end{equation}
After inflation the field $\phi$ oscillates. If one keeps the notation 
$\phi$  for the amplitude of oscillations of this field, then one can say 
that the energy density of this field is given by $\rho(\phi) = {M^2\over 
2}\phi^2$.

To simplify our notation, we will take the scale factor at the end of 
inflation to be $a_0 = 1$. The amplitude of the oscillating field in the 
theory with the potential (\ref{pot}) changes as 
\begin{equation}\label{pot2}
\phi(t) = \phi_0 ~a^{-3/2}(t). 
\end{equation}

The field $\chi$ does not oscillate and almost does not change its 
magnitude until the moment $t_1$ when $H^2(t)= 
 {\rho_\phi\over 3 M_p^2}$ becomes smaller than $m^2/3$.  At that
time one has 
\begin{equation}\label{pot3}
{\rho_\chi\over \rho_\phi} \sim {m^2\chi_0^2\over 6 H^2(t) M_p^2} \sim 
{\chi_0^2\over 2 M_p^2} 
\end{equation}
This ratio, which can also be obtained by a  numerical investigation of 
oscillations of the moduli fields, does not change until the time $t_R$ 
when reheating occurs because $\rho_\chi$ and $\rho_\phi$ decrease in the 
same way: they are proportional to $a^{-3}$. 

At the moment of reheating one has $\rho_\phi(t_R)= \pi^2N(T)T_R^4/30$, 
and the entropy of produced particles $s =2\pi^2 N(T) T_R^3/45$, where 
$N(T)$ is the number of light degrees of freedom. This yields 
\begin{equation}\label{pot4a}
{n_\chi\over s} \sim {\rho_\chi\over m  s} \sim   {\chi_0^2 T_R\over 3m 
M_p^2} 
\end{equation}

Usually one expects $T_R \gg m \sim 10^2$ GeV. Then in order to have 
${n_\chi\over s} < 10^{-12}$ one would need $\chi_0 \ll 10^{-6} M_p$. 
However, it is hard to imagine why the  value of the moduli field at the 
end of inflation should be so small. If one takes $\chi_0 \sim M_p$, 
which looks natural, then one violates the bound ${n_\chi\over s} < 
10^{-12}$ by more than 12 orders of magnitude. This is the essence of the 
cosmological moduli problem \cite{polonyi}.

In general, the situation is more complex. During the expansion of the 
universe the effective potential of the moduli acquires some corrections. 
In particular, quite often the effective mass of the moduli (the second 
derivative of the effective potential) can be represented as 
\begin{equation}\label{mmass}
m^2_\chi  = m^2 +c^2 H^2 \ , 
\end{equation}
where $c$ is some constant and H is the Hubble parameter 
\cite{dinefisch}. Higher derivatives of the effective potential may 
acquire corrections as well. This leads to a different version of the 
moduli problem discussed in \cite{GLV}, see also \cite{modmass}.  The 
position of the minimum of the effective potential of the moduli field in 
the early universe may occur at a large distance from the position of the 
minimum at present. This may fix the initial position of the field $\chi$ 
and lead to its subsequent oscillations. 

A simple toy model illustrating this possibility was given in \cite{LS}: 
\begin{equation}
V = \frac{1}{2} m_\chi^2 \chi^2 
        + \frac{c^2}{2} H^2 \left( \chi - \chi_0 \right)^2 \ .
\label{dmod} 
\end{equation}
At large $H$ the minimum appears at $\chi = \chi_0$; at small $H$ the 
minimum is at $\chi = 0$. Thus one would expect that initially the field 
should stay at $\chi_0$, and later, when $H$ decreases, it  should 
oscillate about $\chi = 0$ with an initial amplitude approximately equal 
to $\chi_0$. The only natural value for $\chi_0$ in supergravity is 
$\chi_0 \sim M_p$. This may lead  to a strong violation of the bound 
(\ref{constr}). 

A more detailed investigation of this situation has shown \cite{LinMod} 
that one should distinguish between three different possibilities: $c\gg 
1$, $c \sim 1$ and $c \ll 1$. 

If $c > O(10)$, the field $\chi$ is trapped in the (moving) minimum of 
the effective potential, its oscillations have very small amplitudes, and 
the moduli problem does not appear at all \cite{LinMod}. This is the 
simplest resolution of the problem, but it is not simple to find 
realistic models where the condition $c > O(10)$ is satisfied. 

The most natural case is $c \sim 1$. It requires a complete study of the 
behavior of the effective potential in an expanding universe. There may 
exist some cases where the minimum of the effective potential does not 
move in this regime, but in general the effects of quantum creation of 
moduli in this scenario \cite{GLV,GRT} are subdominant with respect to 
the classical moduli problem discussed above \cite{GLV,modmass}, so we 
will not discuss this regime in our paper. 

Here we will study the case $c\ll 1$. In this case the effective mass of 
the moduli at $H \gg m$ is always much smaller than $H$, so the field 
does not move towards its minimum, regardless of its position. Thus if 
there is any classical field $\chi_0$ it simply stays at its initial 
position until $H$ becomes smaller than $m$, just as in the case 
considered above, and the resulting ratio ${n_\chi\over s}$ is given by 
Eq. (\ref{pot4a}). 

The moduli problem in this scenario has two aspects. First of all, in 
order to avoid the classical moduli problem one needs to explain why 
$\chi_0 \sim 10^{-6} \sqrt{m\over T_R}\, M_p$, which is necessary (but 
not sufficient) to have ${n_\chi/ s}< 10^{-12}$. Then one should study 
quantum creation of moduli in an expanding universe and check whether 
their contribution to $n_\chi$ violates the bound ${n_\chi/ s}< 
10^{-12}$. This last aspect of the moduli problem was studied in 
\cite{GLV,GRT}. 

In inflationary cosmology   these two contributions (the contributions to 
$n_\chi$ from the classical field $\chi$ and from its quantum 
fluctuations) are almost indistinguishable. Indeed, the dominant 
contribution to the number of moduli produced in an expanding universe is 
given by the fluctuations   of the moduli field produced during 
inflation. These fluctuations have exponentially large wavelengths and 
for all practical purposes they have the same consequences as a 
homogeneous classical field of amplitude  $\chi_0 
 = \sqrt{\langle \chi^2 \rangle}$.

To be more accurate, these fluctuations behave in the same way as the 
homogeneous classical field $\phi$ only if their wavelength is greater 
than $H^{-1}$. During inflation this condition is satisfied for all 
inflationary fluctuations, but after inflation the size of the horizon 
grows and eventually becomes larger than the wavelength of some of the 
modes.  Then these modes begin to oscillate and their amplitude   begins 
to decrease even if at that stage $m < H$. To take this effect into 
account one may simply exclude from consideration those modes whose 
wavelengths become smaller than $H^{-1}$ prior to the moment $t \sim 
m^{-1}$ when $H$ drops down to $m$. It can be shown that in the context 
of our problem this is a relatively minor correction, so we can use the 
simple estimate $\chi_0  = \sqrt{\langle \chi^2 \rangle}$. 

In order to evaluate this quantity we will assume that $c \ll 1$ and $m 
\ll H$ during and after inflation. This reduces the problem to the 
investigation of the production of massless (or nearly massless) 
particles during and after inflation. In the next section we study this 
issue and show that in the most interesting cases where inflationary 
long-wavelength fluctuations of a scalar field can be generated during 
inflation, they give the dominant contribution to particle production. 
This allows us to reduce a complicated problem of gravitational particle 
production to a simple problem which can be easily solved analytically.

\section{Generation of light particles from  and after inflation }

In this section we will present the results of a numerical study of the 
gravitational creation of light scalar particles in the context of 
inflation. Consider a scalar field $\chi$ with the potential 
\begin{equation}
V(\chi) = {1 \over 2}\left(m^2 - \xi R\right) \chi^2 
\end{equation}
where $R$ is the Ricci scalar. In a Friedmann universe $R = -{6 \over 
a^2} (\ddot{a} a + \dot{a}^2)$. The scalar field operator can be 
represented in the form 
\begin{equation}
\chi(x,t) = {1 \over (2 \pi)^{-3/2}} \int d^3k \left[{\hat a}_k \chi_k(t) 
e^{i k x} + {\hat a}^{\dag}_k \chi_k^*(t) e^{-i k x}\right] 
\end{equation}
where the eigenmode functions $\chi_k$ satisfy 
\begin{equation}\label{physmodes}
\ddot{\chi_k} + 3 {\dot{a} \over a} \dot{\chi_k} + \left[\left({k \over 
a}\right)^2 + m^2 - \xi R\right] \chi_k = 0, 
\end{equation}
By introducing conformal time and field variables defined as $\eta \equiv 
\int {dt \over a}, f_k \equiv a \chi_k$ eq. (\ref{physmodes}) can be 
simplified to 
\begin{equation}\label{confmodes}
f_k'' + \omega_k^2 f_k = 0 
\end{equation}
where primes denote differentiation with respect to $\eta$ and 
\begin{equation}
\omega_k^2 = k^2 + a^2 m^2 + \left({1 \over 6} - \xi\right){a'' \over a} 
. 
\end{equation}
The growth of the scale factor is determined by the evolution of the 
inflaton field $\phi$ with potential $V(\phi)$. In conformal time 
\begin{equation}
a'' = {a'^2 \over a} - {8 \pi a\over 3} \left(\phi'^2 - a^2 
V(\phi)\right) 
\end{equation}
\begin{equation}
\phi'' + 2{a' \over a} \phi' + a^2 {\partial V(\phi) \over 
\partial \phi} = 0.
\end{equation}

For initial conditions for the modes $f_k$, in the first approximation
one can use positive frequency vacuum fluctuations $f_k = {1 \over 
\sqrt{2 k}} e^{-i k t}$, see e.g. \cite{tk}. However, when describing 
fluctuations produced at the last stages of a long inflationary period, 
one should begin with fluctuations which have been generated during the 
previous stage of inflation. For example, for massless scalar fields 
minimally coupled to gravity instead of $f_k = {1 \over \sqrt{2 k}} e^{-i 
k t}$ one should use Hankel functions \cite{book}: 
\begin{eqnarray}
f_k(t) =\frac{ia(t)\,{\rm H}}{k\,\sqrt{2\,k}}\, \left(1+\frac{k}{i\,{\rm 
H}}e^{-{\rm H}\,t}\right)\, \exp\left(\frac{i\,k}{{\rm H}}\,e^{-{\rm 
H}\,t}\right), \label{bookeq} 
\end{eqnarray}
where $H$ is the Hubble constant at the beginning of calculations.  To 
make the calculations even more accurate, one should take into account that 
long-wavelength perturbations were produced at early stages of inflation
when $H$ was greater than at the beginning of the calculations.  If 
the stage of inflation is very long, then the final results do not change 
much if instead of the Hankel functions (\ref{bookeq}) one uses $f_k = {1 
\over \sqrt{2 k}} e^{-i k t}$. However, if the inflationary stage is  short, 
then using the functions $f_k = {1 \over \sqrt{2 k}} e^{-i k t}$ 
considerably  underestimates the resulting value of $\langle \chi^2 
\rangle$.

At late times the solutions to Eq. (\ref{confmodes}) can be represented 
in terms of WKB solutions as 
\begin{equation}
f_k(\eta) = {\alpha_k(\eta)\over \sqrt{2\omega_k}} e^{- i\int^{\eta} 
\omega_k d\eta} + {\beta_k(\eta)\over \sqrt {2\omega_k}} e^{+ 
i\int^{\eta} \omega_k d\eta} \ , \label{wkb} 
\end{equation}
where $\alpha_k(\eta)$ and $\beta_k(\eta)$ play the role of coefficients 
of a Bogolyubov transformation. This form is often used to discuss 
particle production because the number density of particles in a given 
mode is given by $n_k=\vert \beta_k(\eta)\vert^2$ and their energy 
density is $\omega_k n_k$. As we will see, though, the main contribution 
to the number density of $\chi$ particles at late times comes from 
long-wavelength modes which are far outside the horizon during reheating. 
As long as they remain outside the horizon these modes do not manifest 
particle-like behaviour, i.e. the mode functions do not oscillate. In 
this situation the coefficients $\alpha$ and $\beta$ have no clear 
physical meaning. We therefore present our results in terms of the mode 
amplitudes $\vert f_k(\eta)\vert^2$, which as we will show contain all 
the information relevant to number density and energy density at late 
times. 

At late times when ${a'' \over a} \sim H < m$ the long wavelength modes 
of $\chi$ will be nonrelativistic and their number density will simply be 
given by Eq. (\ref{constra}). Moreover the very long wavelength modes 
which are still outside the horizon at late times (e.g. at 
nucleosynthesis) will act like a classical homogeneous field whose 
amplitude is given by 
\begin{equation}
\langle \chi^2 \rangle = {1 \over 2 \pi^2 a^2} \int dk k^2 \vert 
f_k\vert^2. 
\end{equation}
It is these very long wavelength modes which will dominate and therefore 
the quantity of interest for us is the amplitude of these fluctuations. 

In our calculations we assumed that $m^2 = c^2 H^2$ with $c \ll 1$; the 
results shown are for $c=0$ but we also did the calculations with $c=.01$ 
and found that the results were independent of $c$ in this range. 

\begin{figure}
\leavevmode\epsfxsize=1\columnwidth \epsfbox{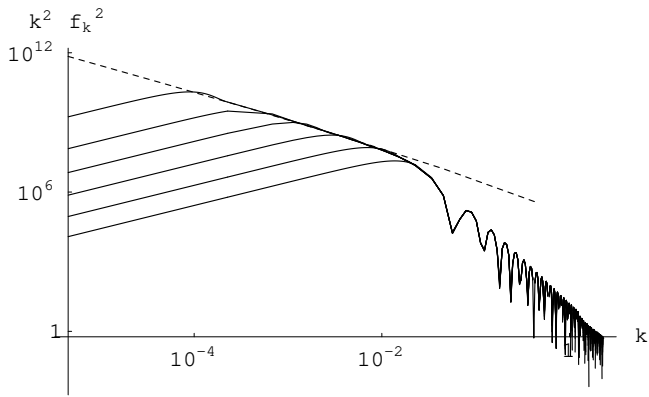} 

\

\caption{Fluctuations vs. mode frequency at a late time. The lower plots 
show runs which started close to the end of inflation, with initial 
conditions $f_k = {1 \over \sqrt{2 k}} e^{-i k t}$. The starting times 
range from $\phi_0=1.5 M_p$ (lowest curve) up through $\phi_0=2 M_p$. The 
highest curve shows the Hankel function solutions given by eqs. 
(\ref{hankel}) and (\ref{correctedhubble})  As we see, calculations with 
$f_k = {1 \over \sqrt{2 k}} e^{-i k t}$ produce the correct spectrum at 
large $k$ but underestimate the level of quantum fluctuations at small 
$k$.} \label{chi2vsp} 
\end{figure}

Figure \ref{chi2vsp} shows the results of solving Eq. (\ref{confmodes}) 
for a model with the inflaton potential $V(\phi) = {1 \over 4} \lambda 
\phi^4$. These data were taken after ten oscillations of the inflaton 
field. The vertical axis shows $k^2 \vert f_k\vert^2$ as a function of 
the momentum $k$. The momentum is shown in units of the Hubble constant 
at the end of inflation. 

The different plots represent runs with different starts of inflation, 
i.e. with different initial values of $\phi$. They all coincide in the 
ultraviolet part of the spectrum, but the runs which started towards the 
end of inflation show a significant suppression in the infrared. This 
shows that fluctuations produced during inflation are the primary source 
of the infrared modes, which in turn dominate the number density. 

The curve on top shows the Hankel function solutions (\ref{bookeq}), 
which give 
\begin{equation}\label{hankel}
\vert f_k\vert^2 = {a^2 H^2 \over 2 k^3} 
\end{equation}
for de Sitter space, i.e. for a constant $H$. In the figure, we have 
corrected this expression by using for each mode the value of the Hubble 
constant at the moment when that mode crossed the horizon. For the 
$\lambda \phi^4$ model shown here the appropriate Hubble parameter for 
each mode can be approximated as 
\begin{equation}\label{correctedhubble}
H_k = \sqrt{2 \pi \lambda \over 3} \left(\phi_e^2 - {1 \over \pi} 
ln\left({k \over H_e}\right)\right) 
\end{equation}
where $\phi_e$ and $H_e$ are the values of the inflaton and the Hubble 
parameter respectively at the end of inflation.

 Note 
that if the Hankel function solutions (\ref{bookeq}) are used as initial 
conditions for a numerical run then they do not change as the modes cross 
the horizon, so the upper curve of the plot can also be obtained from 
such a run. Relative to this upper curve it's easy to see how the 
numerical runs show suppression in the infrared due to starting inflation 
at late times and choosing the initial conditions in the form $f_k = {1 
\over \sqrt{2 k}} e^{-i k t}$, and suppression in the ultraviolet due to 
the end of inflation. The latter suppression is physically realistic. The 
infrared suppression should occur at a wavelength corresponding to the 
Hubble radius at the beginning of inflation. 

The different regions of the graph illustrate effects which occurred at 
different times. During inflation long wavelength modes crossed the 
horizon at early times. Thus the far left portion of the plot shows the 
modes which crossed earliest. They have the highest amplitudes both 
because they were frozen in earliest and because the Hubble constant was 
higher at earlier times when they were produced. The lower plots don't 
show these high amplitudes because inflation began too late for these 
modes to cross outside the horizon and be amplified. Farther to the right 
the curve shows modes which were only slightly if at all amplified during 
inflation. The far right modes were produced during the fast rolling and 
oscillatory stages. These modes are not frozen and can be described 
meaningfully in terms of $\alpha$ and $\beta$ coefficients. The 
regularized expression 
\begin{equation}
\vert f_k\vert^2 = {1 \over \omega_k} \left[\vert \beta_k\vert^2 + 
Re\left(\alpha_k \beta_k^* e^{-2 i \int \omega_k d\eta}\right)\right] 
\end{equation}
shows why the amplitudes of these modes oscillate as a function of $k$. 

In short inflationary fluctuations are primarily responsible for 
producing infrared modes and post-inflationary effects account for 
ultraviolet modes, but it is the infrared modes that were outside the 
horizon at the end of inflation which dominate the number density at late 
times. The earlier inflation began the farther this distribution will 
extend into the infrared, and the long wavelength end of this spectrum 
will always give the greatest contribution to the number density of 
$\chi$ particles. 

Our numerical calculations  are  similar to those of 
 Kuzmin and Tkachev \cite{tk}.
However, they took a rather small initial value of the classical scalar 
field $\phi$, which resulted in less than 60 e-folds of inflation. As 
initial conditions for the fluctuations they used $f_k = {1 \over \sqrt{2 
k}} e^{-i k t}$. They pointed out that the results of such calculations 
can give only a lower bound on the number of $\chi$ particles produced 
during inflation. Consequently, similar calculations performed in 
\cite{GRT} could give only a lower bound on the number of moduli fields 
produced in the early universe. 

Our goal is to find not a lower bound but the complete number of 
particles produced at the last stages of inflation in realistic 
inflationary models, where the total duration of inflation typically is 
much greater than 60 e-foldings. One result revealed by our calculations 
is that the effects of an arbitrarily long stage of inflation can be 
mimicked by the correct choice of ``initial'' conditions chosen for the 
modes $\chi_k$ after inflation. Instead of using the Minkowski space 
fluctuations $f_k = {1\over \sqrt{k}} e^{-ikt}$ used in \cite{tk} as well 
as in our numerical calculations, one should use the de Sitter space 
solutions (\ref{bookeq}), with $H$ corrected to the value it had for each 
mode at horizon crossing. Using these modes at the end of inflation is 
equivalent to running a simulation with a long stage of inflation.  

Our numerical calculations confirmed the result  that we are going to 
derive analytically in the next section (see  also \cite{NO}): The number 
of $\chi$ particles ($n_\chi \sim m\langle  \chi^2\rangle$) produced 
during the stage of inflation beginning at  $\phi = \phi_0$ in the 
simplest model $M^2\phi^2/2$ is proportional to  $\phi_0^4$, whereas in 
the model $\lambda\phi^4/4$ it is proportional  to $\phi_0^6$. Thus the 
total number of particles produced during  inflation is {\it extremely} 
sensitive to the choice of initial  conditions. If one considers $\phi_0$ 
corresponding to the beginning  of the last 60 e-folds of inflation, the 
total number of particles  produced at that stage appears to be much 
greater than the lower bound  obtained in \cite{tk}. As we will see, this 
will allow us to put a  much stronger constraint on the moduli theories 
than the constraint  obtained in \cite{GRT}. 

\section {Light moduli from  inflation }

The numerical results obtained in the previous section confirm our 
expectation that in the most interesting cases where long-wavelength 
inflationary  fluctuations of light scalar fields can be generated during 
inflation, they give the dominant contribution to particle production. In 
particular, in the case of $c \ll 1$, $m \ll H$  most moduli field 
fluctuations are generated during inflation rather than during the 
post-inflationary stage. These fluctuations grow  at the stage of 
inflation  in the same way as if the moduli field $\chi$ were massless 
\cite{book}: 
\begin{equation}\label{4a}
{d\langle \chi^2 \rangle\over dt}  = {H^3\over 4\pi^2} . 
\end{equation}
If the Hubble constant does not change during inflation, one obtains the 
well-known relation 
\begin{equation}\label{5aa}
{ \langle \chi^2 \rangle }  = {H^3\, t\over 4\pi^2}. 
\end{equation}
However, in realistic inflationary models the Hubble constant changes in 
time, and fluctuations of the light fields $\chi$ with $m \ll H$ behave 
in a more complicated way. 

As an example, let us consider the case studied in the last section. Here 
inflation is driven by a field $\phi$ with an effective potential 
$V(\phi) = {\lambda\over 4}\phi^4$ at $\phi > 0$. This potential could be 
oscillatory or flat for $\phi < 0$. We consider a light scalar field 
$\chi$ which is not coupled to the inflaton field $\phi$, and which is 
minimally coupled to gravity. 

The field $\phi$ during inflation obeys the following equation: 
\begin{equation}\label{1a}
3H\dot\phi= - \lambda \phi^3  . 
\end{equation}
Here 
\begin{equation}\label{2a}
H= {1\over 2}\sqrt{\lambda\over 3}~ {\phi^2\over M_p}  . 
\end{equation}
These two equations yield the solution \cite{book} 
\begin{equation}\label{3a}
\phi =  \phi_0 ~\exp\left(-2\sqrt{\lambda\over 3} M_p t\right), 
\end{equation}
where $\phi_0$ is the initial value of the inflaton field $\phi$. In this 
case Eq. (\ref{4a}) reads: 
\begin{equation}\label{5a}
{d\langle \chi^2 \rangle\over dt}  = {\lambda \sqrt \lambda\over 96\sqrt 
3 \pi^2  } {\phi_0^6 \over M_p^3} \exp\left(-12\sqrt{\lambda\over 3} M_p t\right)
\end{equation}

The result of integration at large $t$ converges to 
\begin{equation}\label{6a}
{ \langle \chi^2 \rangle }  = 
  {\lambda\, \over 2}\left({ \phi_0^3\over 24\pi M_p^2}\right)^2 .
\end{equation}
This result agrees with the results of our numerical investigation 
described in the previous section. 

>From the point of view of the moduli problem, these fluctuations
lead to the same consequences as a classical scalar field $\chi$ which is 
homogeneous on the scale $H^{-1}$ and which has a typical amplitude 
\begin{equation}\label{7a}
 \chi_0 = \sqrt{\langle \chi^2 \rangle}  = \sqrt{\lambda
\over 2}~{\phi_0^3\over 24\pi M_p^2} . 
\end{equation}

 A similar result can be obtained in the model $V(\phi)  = {M^2
\over 2} \phi^2$. In this case one has \cite{book} 
\begin{equation}\label{8aa}
 \phi(t) = \phi_0-  \sqrt {2\over   3 }\,   M_p Mt .
\end{equation}
The time-dependent Hubble parameter is given by 
\begin{equation}\label{8aaa}
H = {M\over \sqrt 6 M_p}\phi(t)  , 
\end{equation}
which yields 
\begin{equation}\label{8aaaa}
 \chi_0 = \sqrt{ \langle \chi^2 \rangle }  =   {M
\phi_0^2\over 8\pi\sqrt {3} M_p^2 } . 
\end{equation}

As we see, the value of $\chi_0$ depends on the initial value of the 
field $\phi$. This result has the following interpretation. One may 
consider an inflationary domain of initial size $H^{-1}(\phi_0)$. This 
domain after inflation becomes exponentially large. For example, its size 
in the model with $V(\phi)  = {M^2\over 2} \phi^2$ becomes \cite{book} 
\begin{equation}\label{8aab}
l \sim H^{-1}(\phi_0)\exp\left({\phi_0^2\over 4M_p^2}\right)  . 
\end{equation}
In order to achieve 60 e-folds of inflation in this model one needs to 
take $\phi_0 \sim 15 M_p$. This implies that a typical value of the 
(nearly) homogeneous scalar field $\chi$ in a universe which experienced 
60 e-folds of inflation in this model is given by 
\begin{equation}\label{8aau}
 \chi_0 = \sqrt{ \langle \chi^2 \rangle } \sim
5M . 
\end{equation}
In realistic versions of this model one has $M \sim   5\times 10^{-6} M_p 
\sim  10^{13}$ GeV \cite{book}. Substitution of this result into Eq. 
(\ref{pot4a}) gives 
\begin{equation}\label{pot4aaa}
{n_\chi\over s} \sim  2\times  10^{-10}~ {T_R  \over  m}\ . 
\end{equation}
This implies that the condition ${n_\chi/ s} \lesssim 10^{-12}$ requires 
that the reheating temperature in this model should be at least two 
orders of magnitude smaller than $m$. For example, for $m \sim 10^2$ GeV 
one should have $T_R \lesssim 1$ GeV, which looks rather problematic. 

This result confirms the basic conclusion of Ref. \cite{GLV} that the 
usual models of inflation do  not solve the moduli problem. Our result is 
similar to the result obtained in \cite{GRT} by numerical methods, but it 
is approximately  two orders of magnitude stronger. The reason for this 
difference is that the authors of Ref. \cite{GRT} used a much smaller 
value of $\phi_0$ in their numerical calculations. Consequently, they 
took into account only the particles produced at the very end of 
inflation, whereas the leading effect occurs at earlier stages of 
inflation, i.e. at larger $\phi$. 

In  general  one can get a simple estimate of $\chi_0 = \sqrt{ \langle 
\chi^2 \rangle }$ by assuming that the universe expanded with a constant 
Hubble parameter $H_0$ during the last 60 e-folds of inflation. To make 
this estimate more accurate one should take the value of the Hubble 
constant not at the end of inflation but approximately 60 e-foldings 
before it, at the time when the fluctuations on the scale of the present 
horizon were produced. The reason is that the largest contribution to the 
fluctuations is given by the time when the Hubble constant took its 
greatest values. Also, at that stage the rate of change of H was 
relatively slow, so the approximation $H = H_0 =const$ is reasonable. 
Thus one can write 
\begin{equation}\label{8aauu}
 \chi_0 = \sqrt{ \langle \chi^2 \rangle } \gtrsim
{H_0   \over 2\pi }\sqrt{H_0 t} \sim {H_0   \over 2\pi }\sqrt{60} \sim 
H_0 \ . 
\end{equation}
This gives 
\begin{equation}\label{pot4au}
{n_\chi\over s} \gtrsim   {T_R \,H_0^2 \over 3m\, M_p^2} \ . 
\end{equation}
In the simplest versions of chaotic inflation with potentials 
$M^2\phi^2/2$ or $\lambda\phi^4/4$ one has  $H_0 \sim 10^{14}$ GeV, which 
leads to the requirement $T_R \lesssim 1$ GeV. But this equation shows 
that there is another way to relax the problem of the gravitational 
moduli production: one may   consider models of inflation with a very 
small value of $H_0$ \cite{GLV}. For example, one may have ${n_\chi/ s} 
\sim 10^{-12}$ for $T_R \sim H_0 \sim 10^7$ GeV. 

However, this condition is  not sufficient to resolve the moduli problem; 
the situation is more complicated. First of all, it is very difficult to 
find inflationary models where inflation occurs only during 60 
e-foldings. Typically it lasts much longer, and the fluctuations of the 
light moduli fields will be much greater. This is especially obvious in 
the theory of eternal inflation where the amplitude of fluctuations of 
the light moduli fields can become indefinitely large \cite{book}. In 
particular, if the condition $m^2_\chi \sim m^2 +c^2 H^2$ with $c \ll 1$ 
remains valid for $\chi \gtrsim M_p$, then  one may  expect the 
generation of moduli fields $\chi > M_p$. This should initiate inflation  
driven  by the light moduli \cite{NO}. Then the situation would become 
even more problematic:  we would need to find out how one could produce 
baryon asymmetry of the universe after the light moduli decay and how one 
could obtain  density perturbations $\delta\rho/\rho \sim 10^{-4}$ in 
such a low-scale inflationary model. 

One may expect that the region of values of $\chi$ where its effective 
potential has small curvature $m^2 \ll H^2$ may be limited, and may even 
depend on $H$. Then the existence of a long stage of inflation would push 
the fluctuations of the field $\chi$ up to the region where its effective 
potential becomes curved, and instead of our results for $\chi_0$ one 
should substitute the largest value of $\chi$ for which $m^2_\chi < H^2$. 
In such a situation one would have  a mixture of problems which occur at 
$c \ll 1$  and at $c \sim 1$.

Finally, we should emphasize that all our worries about quantum creation 
of moduli appear only after one makes the assumption that for whatever 
reason the initial value of the classical field $\chi$ in the universe 
vanishes, i.e. that the classical version of the moduli problem has been 
resolved. We do not see any justification for this assumption in theories 
where the mass of the moduli field in the early universe is much smaller 
than $H$. Indeed, in such theories the classical field $\chi$ initially 
can take {\it any} value, and this value is not going to change until the 
moment $t \sim H^{-1} \sim m^{-1}$. The main purpose of this paper was to 
demonstrate that even if one finds a solution to the light moduli problem 
at the classical level, the same problem will appear again because of 
quantum effects. 

This does not mean that the moduli problem is unsolvable. One of the most 
interesting solutions  is provided by thermal inflation \cite{LS}. The 
Hubble constant during inflation in this scenario is very small, and the 
effects of moduli production there are rather insignificant. Another 
possibility is that moduli are very heavy in the early universe, 
$m^2_{\chi} = m^2 + c^2 H^2$, with $c 
> O(10)$, in which case the moduli problem does not appear
\cite{LinMod}. The main question is whether we really need to make the 
theory so complicated   in order to avoid the cosmological problems 
associated with moduli. Is it possible to find a simpler solution? One of 
the main results of our investigation is to confirm the conclusion of 
Ref. \cite{GLV} that the simplest versions of inflationary theory do not 
help us to solve the moduli problem but rather aggravate it. 

In conclusion we would like to note that the methods developed in this 
paper apply not only to the theory of moduli production but to other 
problems as well. For example, one may study the theory of gravitational 
production of superheavy scalar particles after inflation \cite{ch,tk}. 
If these particles are minimally coupled to gravity and have mass $m \ll 
H$ during inflation, then one can use our Eqs. (\ref{7a}), (\ref{8aaaa}) 
to calculate the number of produced particles. These equations imply that 
the final result will strongly depend on $\phi_0$, i.e. on the duration 
of inflation. If inflation occurs for more than 60 Hubble times, the 
production of  particles with $m \ll H$ is much more efficient than was 
previously anticipated. As we just mentioned, if $\phi_0$ is large enough 
then the production of fluctuations of the field $\chi$ may even lead to 
a new stage of inflation driven by the field $\chi$ \cite{NO}. On the 
other hand, if $m$ is greater than the value of the Hubble constant at 
the very end of inflation, then quantum fluctuations are  produced only 
at the early stages of inflation (when $H > m$). These fluctuations 
oscillate and decrease exponentially during the last stages of inflation. 
In such cases the final number of  produced particles will not depend on 
the duration of inflation and can be unambiguously calculated. We hope to 
return to this question in a separate publication. 

The authors are grateful to I. Tkachev for useful discussions. This work 
was supported  by  NSERC and  CIAR and by NSF grant AST95-29-225. The 
work of G.F. and A.L. was also supported   by NSF grant PHY-9870115, and
the work of L.K. and A.L. by NATO Linkage Grant 975389.

\end{document}